\documentclass[aps,prc,twocolumn,showpacs,amsmath,amssymb,nofootinbib]{revtex4}

\usepackage{color} 
\usepackage{epsfig}

\begin{document}
\title{\boldmath Nuclear matter with JISP16 $NN$ interaction}
\author{A. M. Shirokov}
\affiliation{Skobeltsyn Institute of Nuclear Physics, Moscow State
University, Moscow 119991, Russia}
\affiliation{Department of Physics and Astronomy, Iowa State University, Ames, IA 50011, USA}
\affiliation{Pacific National University, 136
Tikhookeanskaya Street, Khabarovsk 680035, Russia}
\author{A. G. Negoita}
\affiliation{Department of Physics and Astronomy, Iowa State University, Ames, IA 50011, USA}
\author{J. P. Vary}
\affiliation{Department of Physics and Astronomy, Iowa State University, Ames, IA 50011, USA}
\author{S. K. Bogner}
\affiliation{National Superconducting Cyclotron Laboratory and
Department of Physics and Astronomy,  Michigan State
University, East Lansing, MI 48844, USA}
\author{A. I. Mazur}
\affiliation{Pacific National University, 136
Tikhookeanskaya Street, Khabarovsk 680035, Russia}
\author{E. A. Mazur}
\affiliation{Pacific National University, 136
Tikhookeanskaya Street, Khabarovsk 680035, Russia}
\author{D. Gogny}
\affiliation{Lawrence Livermore National Laboratory, Livermore, CA
94551, USA}

\begin{abstract}
Saturation properties of the JISP16 $NN$ interaction are studied in
symmetric nuclear matter calculations, with special attention paid to the convergence properties with respect to the number of partial waves. We also present results of pure neutron matter calculations with the JISP16 interaction. 
\end{abstract}

\maketitle

Conventional nuclear matter corresponds to the infinite Coulomb-free
system with the same number of protons and neutrons and uniform
density. Such an idealized model, whose properties are inferred by extrapolating from known nuclei,  is a useful tool for studying saturation
properties of inter-nucleon forces. In this paper we calculate nuclear
matter properties with the JISP16 $NN$ interaction.

The JISP16 $NN$ interaction proposed in Refs. \cite{JISP16,JISP16-f} is constructed in
the $J$-matrix inverse scattering approach~\cite{ISTP}. It is known to
provide an excellent description of $np$ scattering data with
$\chi^2/datum\approx 1$~\cite{Machl}. The interaction was fitted in
Ref. \cite{JISP16} by means of phase-equivalent transformations to
the binding energies of nuclei with $A\leq 16$, and it provides a
good description of bindings and spectra of light nuclei without
referring to three-nucleon forces \cite{JISP16,YaF,Rila,extrap08,fb19,14F,Chase,Be, MArisBe, MAris12C, Kristina12C, MAris,rotB, jvaryNTSE13, PieterNTSE13,StrBlokh}. 
In particular, the binding energy and spectrum of exotic
proton-excess nucleus $^{14}$F have been predicted \cite{14F} in No-core Full
Configuration Calculations  \cite{extrap08} with the JISP16 $NN$
interaction. These predictions were confirmed in a subsequent experiment~\cite{14F-ex} where this nucleus was first observed. 

A difficulty in nuclear matter studies with conventional $NN$ interactions
is that the calculations are nonperturbative due to the strong short-range repulsion and tensor forces \cite{Bethe}. However, as it was shown in
Refs. \cite{Scott-05,Hebeler-11, Hebeler-13}, in the case of soft $NN$ interactions, a
perturbative approach can be successfully used for nuclear matter
calculations. In particular, the authors of Refs. \cite{Scott-05,Hebeler-11} 
demonstrated that the dominant particle-particle channel contributions become perturbative 
in nuclear matter calculations using so-called low-momentum $NN$ interactions 
($V_{{\rm low}-k}$) obtained by renormalization group methods
\cite{RG1,RG2} from Argonne AV$_{18}$ and chiral effective field theory
N$^3$LO $NN$ interactions.

JISP16 is a soft $NN$ interaction providing faster convergence of
nuclear structure calculations than typical realistic $NN$ interactions
providing high quality fits to all $NN$ data. The interaction is completely nonlocal:
by construction, it is given by a matrix in the harmonic oscillator
basis in each partial wave of the $NN$ interaction. Therefore there is
nothing like a core in this interaction which is 
a leading source of the
nonperturbative behavior of the nuclear matter calculations. The
structure of the interaction guarantees description of the $NN$
scattering phase shifts up to the energy of 350 MeV in the lab frame. At
the energies of about 400 MeV and higher the JISP16 scattering phase
shifts exponentially drop to zero.  
The $V_{{\rm low}-k}$ $NN$ interactions of Ref. \cite{Scott-05}.
have a similar falloff when renormalized to a corresponding scale 
which lends support to our adoption of the perturbative approach
of Refs. \cite{Scott-05,Hebeler-11}.


Nuclear matter  is known \cite{Scott-05,Hebeler-11,collapse} to collapse with the  
$V_{{\rm low}-k}$ interactions if the renormalization group evolution is truncated at
the two-body level. Saturation is however restored if one includes the corresponding low-momentum three-nucleon interactions that are induced by the renormalization group evolution \cite{Scott-05,Hebeler-11}. See also Ref.~\cite{no-collapse-rel}. In view of the similarities between JISP16 and the $V_{{\rm low}-k}$ $NN$ interactions, one might expect that the soft JISP16 interaction would also fail to produce nuclear matter saturation at the $NN$-only level. It is easy
to show that the JISP-like interactions represented by a matrix in the
oscillator basis 
cause collapse in nuclear matter 
in a pure
Hartree--Fock calculation if 
the trace of the two-body interaction matrix is
negative. The trace of the JISP16 interaction matrix is positive, hence
this $NN$ force does not collapse nuclear matter, at least at the Hartree--Fock level.

\begin{figure}
\epsfig{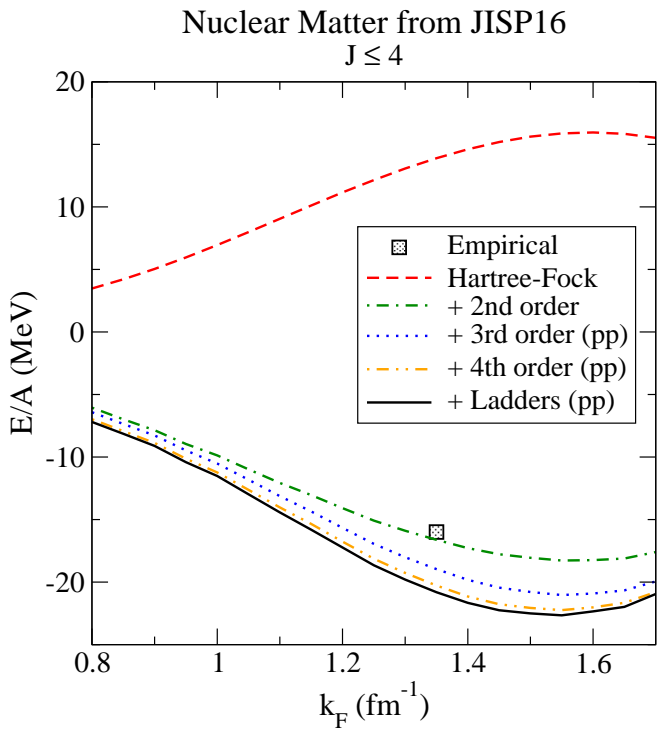}
\caption{(Color online) Results of perturbative nuclear matter 
calculations with JISP16 $NN$ interaction.  }
\label{pert}
\end{figure}

The Hartree--Fock approximation is however very inaccurate as is seen
from Fig. \ref{pert} where we present results obtained in a sequence of 
approximations, including contributions up to total angular momentum $J=4$ in the $NN$ interaction. In particular, we performed calculations in a pure
Hartree--Fock approximation, 
then  including second order corrections,
then  including particle-particle third and fourth order corrections, and
then summing the ladder particle-particle contributions to all orders. In all cases, the single-particle energies are dressed at the Hartree--Fock level, including the full momentum dependence of the Hartree--Fock single particle potential. The Pauli blocking operator is treated in the angle-average approximation, which has been shown to be accurate to $\sim .5$ MeV per nucleon level for soft interactions \cite{Baardsen}.
This sequence is seen to converge. The converged energy minimum corresponds to higher
density (larger Fermi momentum $k_{\rm F}\approx 1.55\rm~fm^{-1}$) 
and larger binding energy ($\approx 22.7$~MeV per nucleon)
than the empirical nuclear matter saturation point.

\begin{figure}
\epsfig{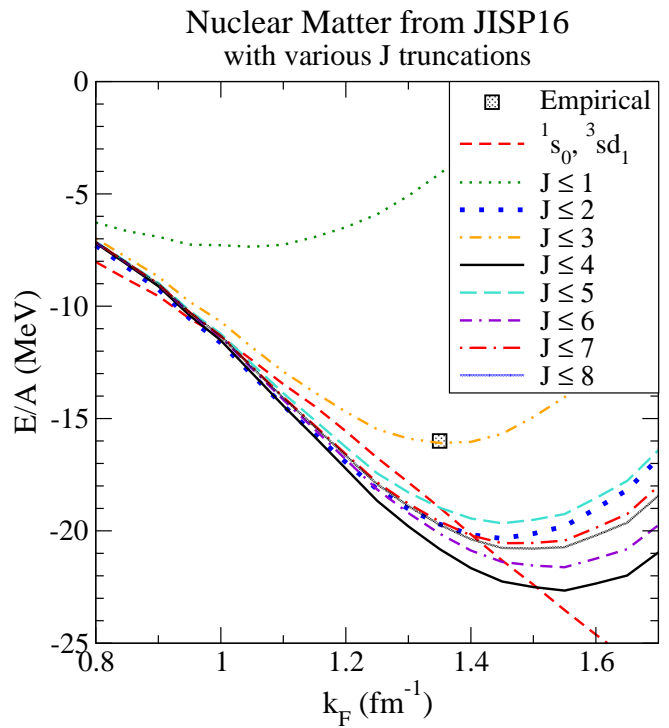}
\caption{(Color online) Convergence of nuclear matter calculations with
respect to the $J$-truncation of the JISP16 $NN$ interaction.  The calculations include particle-particle ladder diagrams to all orders with Hartree--Fock single-particle energies.}
\label{diffJ}
\end{figure}

Figure \ref{diffJ} demonstrates the convergence of the nuclear matter
equation of state when potential energy contributions with increasing
total angular momentum $J$ are successively  
included in the calculations. The conventional JISP16
interaction of Refs. \cite{JISP16,JISP16-f} is defined in $NN$ partial
waves with $J\leq 4$ only; just these $J\leq 4$ results were
presented in Fig. \ref{pert}. It is clear from Fig. \ref{diffJ} 
that interaction in partial waves with $J\leq 4$ is not enough to achieve
convergence  at the higher densities. The sensitivity of nuclear
matter saturation properties to 
higher partial waves was also mentioned in 
Refs. \cite{highJ70,highJ79,highJ10}. 

We extended the JISP16 
interaction to higher partial waves using the $J$-matrix inverse
scattering approach described in detail in Ref. \cite{ISTP} and used Nijmegen
partial waves analysis \cite{Nijm-PWA} as an input. The JISP16
interaction is defined with the truncation in oscillator quanta 
$N=2n+L\leq 9$. Hence this interaction can be defined only in $NN$
partial waves  with $J\leq 8$: the potentials in 
partial waves with orbital momenta $L=8$ and 9 are presented by $1\times 1$
matrices  in the oscillator basis. Nevertheless, the Nijmegen phase shifts are reasonably well
reproduced even in the partial waves with the highest possible angular
momenta. 

The results obtained with this $J$-extended JISP16 $NN$ interaction are
also presented in Fig. \ref{diffJ}. It is seen that the convergence of
the nuclear matter equation of state with
respect to $J$ is achieved when the interaction in all partial waves with
$J\leq 7$ is included in our calculations.

\begin{figure}
\epsfig{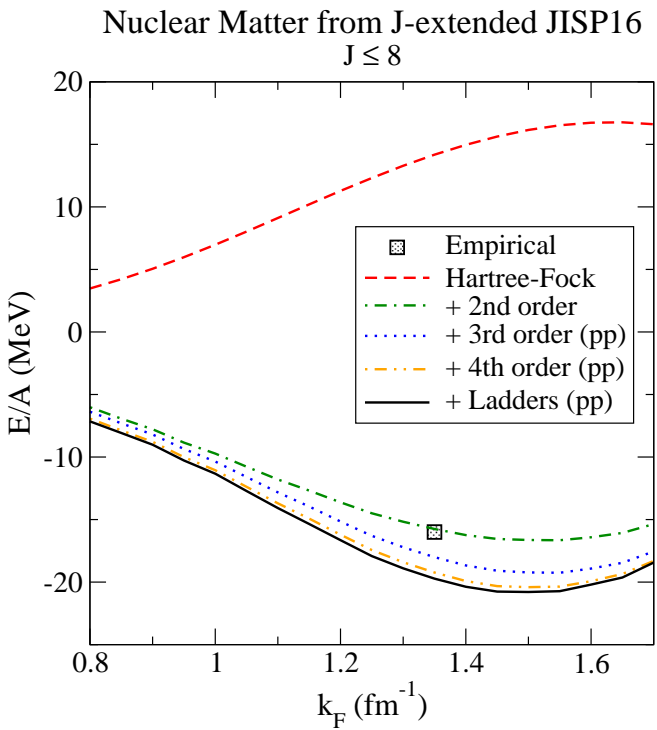}
\caption{(Color online) Results of perturbative nuclear matter 
calculations with JISP16 $NN$ interaction extended up to ${J=8}$.  }
\label{pertJ8}
\end{figure}

The convergence of our sequence of nuclear matter calculations with the 
JISP16 $NN$ interaction extended up to
$J=8$ partial waves is illustrated by Fig. \ref{pertJ8}. The saturation
point in this case is slightly shifted to smaller densities and smaller
binding energies as compared to the results obtained with the
conventional JISP16 interaction (see Fig. \ref{pert}). Nevertheless this
$J$-extended JISP16 interaction still overbinds and overcompresses
nuclear matter. 

\begin{figure}
\epsfig{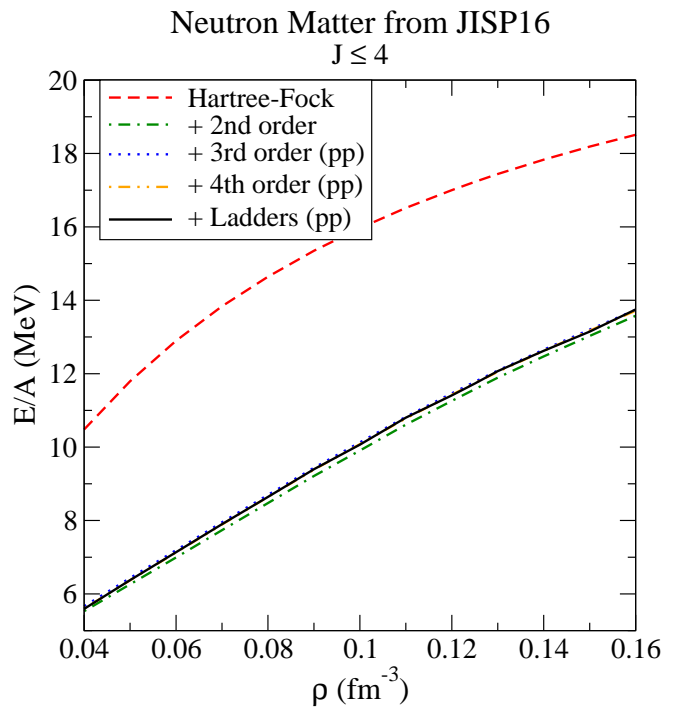}
\caption{(Color online) Results of perturbative pure neutron  matter 
calculations with JISP16 $NN$ interaction.  }
\label{PNMpert}
\end{figure}

\begin{figure}
\epsfig{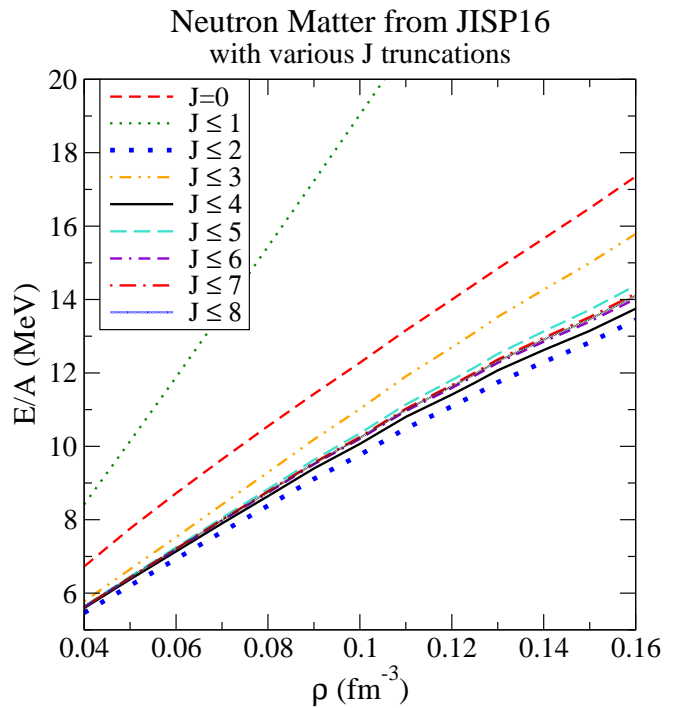}
\caption{(Color online) Convergence of  pure neutron matter  calculations
 allowing for corrections up to the fourth order and summing of ladder
 diagrams  with
respect to the $J$-truncation of the JISP16 $NN$ interaction.  }
\label{PNMdiffJ}
\end{figure}

\begin{figure}
\epsfig{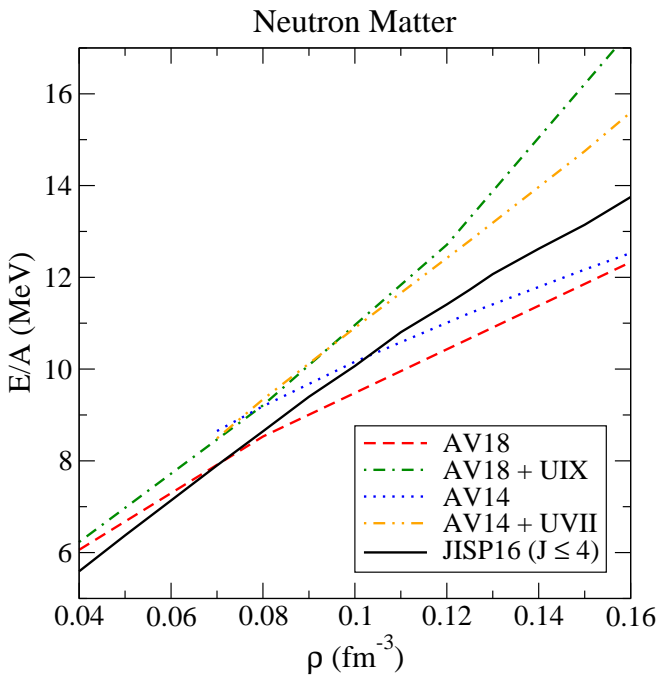}
\caption{(Color online) Comparison of  pure neutron matter equations of state obtained
with JISP16 and other interaction models.  }
\label{PNMcomp}
\end{figure}

It is interesting also to obtain predictions for the pure neutron matter  equation
of state with JISP16. Our perturbative approach is seen from Fig. \ref{PNMpert} to converge.
The convergence of the neutron matter energy with respect to $J$ (see Fig. \ref{PNMdiffJ}) 
is achieved much faster than in
the case of the symmetric nuclear matter. Figure \ref{PNMcomp} presents a 
comparison of the JISP16 induced neutron matter 
equation of state with the results obtained with Argonne AV14 $NN$ interaction solely
and in combination with Urbana UVII $NNN$ force \cite{Wiringa} 
and with Argonne AV18 $NN$ interaction solely
and in combination with Urbana UIX $NNN$ force \cite{Akmal}. It appears  that JISP16
generates pure neutron matter properties at high densities intermediate between predictions
of conventional realistic $NN$ and
$NN+NNN$ interaction models.

Summarizing, we have shown that the soft and nonlocal JISP16 interaction gives a saturating nuclear matter equation of state that converges rapidly in many-body perturbation theory, at least in the particle-particle channel. Due to the soft nature of the JISP16 interaction, and the fact that the dominant contributions to bulk properties of nuclei and nuclear matter are known to be given by the Brueckner Hartree--Fock type correlations treated in our calculations, we expect a more sophisticated many-body treatment will not substantially alter our conclusions, especially for pure neutron matter where it has been shown that low-order perturbative calculations reproduce sophisticated coupled-cluster \cite{Hagen:2013}  and Auxiliary-Field Diffusion Monte Carlo (AFDMC) calculations \cite{Gezerlis:2013, Tews:2013} for sufficiently soft input interactions.  Our symmetric nuclear matter calculations, on the other hand, come with somewhat larger uncertainties due to the omission of particle-hole and three-body correlations in the medium, both of which contribute at the 1 MeV/nucleon level near saturation for coupled-cluster calculations using soft chiral effective field theory $NN$ and $NNN$ interactions \cite{Hagen:2013}.  

The saturation property of the JISP16 potential differs
from $V_{{\rm low}-k}$ $NN$ interactions at low cutoffs (${\Lambda \lesssim 3.0}$~fm$^{-1}$) that give comparably soft and nonlocal potentials, 
which indicates that the adjustment of the JISP16 off-shell properties
by fitting light nuclei may simulate some contributions attributable to 
three-body forces in the $V_{{\rm low}-k}$ approach.
However, the JISP16 saturation point is still overbound at too high a density
as compared to conventional extrapolations to the
infinite mass limit of heavy nuclei properties. The main idea of the
JISP-type interaction is to utilize an {\em ab exitu} approach 
\cite{JISP6,JISP6G,JISP16} in the $NN$ force design, i. e., first the
$J$-matrix inverse scattering approach \cite{ISTP} is used to construct
an $NN$ interaction perfectly describing the two-nucleon data (deuteron
properties and $NN$ scattering), next the interaction is modified by
phase-equivalent transformations in order to achieve a reasonable
description of many-body nuclear systems. Following this route, the
JISP6 interaction fitted to nuclei with $A\leq 6$ was proposed in
Refs. \cite{JISP6,JISP6G}. A subsequent phase-equivalent modification of
this $NN$ interaction resulted in construction of the JISP16 version
\cite{JISP16} fitted to nuclei with $A\leq 16$. The nuclear matter
overbinding presented here poses a challenge to 
develop a subsequent phase-equivalent modification of JISP16 that
achieves an improved 
description of the nuclear matter saturation without sacrificing the good description
of light nuclei. 
Such an improved interaction may also improve the description of
$N\approx Z$ nuclei with $A\geq 12$ --- the overbinding of nuclei at the
end of the $p$ shell 
that was revealed 
with the help of 
very accurate {\em ab initio} NCFC
approach (see review~\cite{StrBlokh})  introduced  in Ref. \cite{extrap08}.

It is interesting to
note that the JISP16 interaction with the $J\leq 3$ truncation provides
a nuclear matter equation of state with the minimum at the
phenomenological saturation point (see Fig. \ref{diffJ}). Higher-$J$ 
interaction terms shift  
 the equation of state minimum. The 
high-$J$ sensitivity of the saturation point can be used to fit
the interaction to the nuclear matter properties. 
In fact, the
$J$-dependence depicted in Fig. \ref{diffJ} suggests 
how to design a set of
phase-equivalent transformations of the JISP16 interaction in the 
$J\geq 4$ partial waves that will result in cancellation of these
high-$J$ interaction terms in the nuclear matter calculations. On the
other hand, nuclei with $A\leq 16$ are insensitive to 
these high-$J$ $NN$ interactions. Therefore the suggested 
fitting procedure should not affect the description of light nuclei involved in
the initial fit of the JISP16 $NN$ interaction.

The fact that a soft NN interaction, such as JISP16 truncated
at $J\leq 3$, provides a reasonable saturation curve for nuclear matter 
is itself an interesting result.  It demonstrates that the long-held belief 
that soft NN interactions cannot properly saturate nuclear matter~\cite{Bethe:1971} 
is not strictly true.

\begin{acknowledgements}

We gratefully acknowledge discussions with Fritz Coester.
This work was supported by the US DOE Grants 
DE-FG02-87ER40371 and DESC0008485 (SciDAC-3/NUCLEI), 
by the US National Science Foundation under Grants No. PHY-0758125, PHY-1068648 and PHY 0904782, 
and by the Ministry of
Education and Science of the Russian Federation under   Contracts
P521 and 14.V37.21.1297.  
\end{acknowledgements}


\begin{thebibliography}{99}
\bibitem{JISP16} A. M. Shirokov, J. P. Vary, A. I. Mazur, and
    T. A. Weber, Phys. Lett. B {\bf 644}, 33 (2007).

\bibitem{JISP16-f} A Fortran code for the JISP16 interaction
  matrix elements is available at http:/\!/nuclear.physics.iastate.edu.

\bibitem{ISTP} A. M. Shirokov, A. I. Mazur, S. A. Zaytsev, J. P. Vary,  
    and T. A. Weber, Phys. Rev. C {\bf 70}, 044005 (2004);
 in {\itshape The $J$-Matrix  Method. Developments and Applications},
        edited by A. D. Alhaidari, H. A. Yamani, E. J. Heller, and
        M. S. Abdelmonem (Springer, 2008), 219.

\bibitem{Machl} R. Machleidt, {\em private communication} (2006). 
The $NN$ data bases and analyses can be obtained also in a web-site
based format by the online SAID facility ({http:/\!/gwdac.phys.gwu.edu}),
R.~A.~Arndt, W.~J.~Briscoe, I.~I.~Strakovsky, and R.~L.~Workman, Phys.\ Rev.\ C\ \textbf{76}, 025209 (2007). 

\bibitem{YaF} A. M. Shirokov, J. P. Vary, A. I. Mazur, and T. A. Weber,
Phys. At. Nucl. {\bf 71}, 1232 (2008).

\bibitem{Rila} A. M. Shirokov, J. P. Vary, and P. Maris,  
in {\em Proc. 27th Int. Workshop on Nucl. Theory,
Rila Mountains, Bulgaria, 23--28 June, 2008}, edited by  S.~Dimitrova
(Bulgarian Acad. Sci., Sofia,  2008), p. 205;
arXiv:0810.1014 [nucl-th].

\bibitem{extrap08} P. Maris, J. P. Vary, and A. M. Shirokov,
 Phys. Rev. C {\bf 79},  014308 (2009).

\bibitem{fb19} A. M. Shirokov, V. A. Kulikov, P. Maris, A. I. Mazur,
E. A. Mazur, and J. P. Vary,  EPJ Web of Conf. {\bf 3} (Proc. 19th Int.
IUPAP Conf. on Few-Body Problems in Phys.), 05015 (2010).


\bibitem{14F}  P. Maris,  A. M. Shirokov, and J. P. Vary,  Phys. Rev. C {\bf 81},  
  021301 (2010).
  
\bibitem{Chase} C.~Cockrell, J.~P. Vary, and P.~Maris, Phys. Rev. C {\bf 86}  (2012)
  034325.

\bibitem{Be} P. Maris, J. Phys. Conf. Ser. {\bf 402}, 012031 (2012).

\bibitem{MArisBe} P. Maris,  J. Phys. Conf. Ser. {\bf 445}, 012035 (2013) .
  
  

\bibitem{MAris12C} P.~Maris, H.~M.~Aktulga, M.~A.~Caprio, U.~V.~\c{C}ataly\"urek, E.~G.~Ng, D.~Oryspayev, H.~Potter, 
E.~Saule, M.~Sosonkina, J.~P.~Vary, C.~Yang, and Z.~Zhou,
J. Phys. Conf. Ser. {\bf 403}, 012019 (2012).

\bibitem{Kristina12C} K.~D.~Launey, T.~Dytrych, J.~P.~Draayer, G.~K.~Tobin, M.~C.~Ferriss, D.~Langr,
A.~C.~Dreyfuss, P.~Maris, J.~P.~Vary, and C.~Bahri, in {\em Fission and Properties of Neutron-Rich Nuclei},
Proc. Fifth Int. Conf. ICFN5. Sanibel Island, 
Florida, USA, 4--10 November 2012. 
World Scientific, 2013, p.~29; 
doi:~10.1142/9789814525435\verb+_+0003.

\bibitem{MAris} P. Maris and J. P. Vary, Int. J. Mod. Phys. E {\bf 22},1330016 (2013) .

\bibitem{rotB} M. A. Caprio, P. Maris, and J. P. Vary, Phys. Lett. B {\bf 719}, 179 (2013).

\bibitem{jvaryNTSE13} J. P. Vary, Proc. Int. Conf. Nucl.  Theor.  Supercomputing Era~--- 2013 (NTSE-2013).
Eds. A.~M.~Shirokov and A.~I.~Mazur. Pacific National University, Khabarovsk, Russia, 2014, p. 15, 
\mbox{http:/\!/www.ntse-2013.khb.ru/Proc/JPVary.pdf}.

\bibitem{PieterNTSE13} P. Maris, Proc. Int. Conf. Nucl.  Theor.  Supercomputing Era~--- 2013 (NTSE-2013).
Eds. A.~M.~Shirokov and A.~I.~Mazur. Pacific National University, Khabarovsk, Russia, 2014, p. 37, 
\mbox{http:/\!/www.ntse-2013.khb.ru/Proc/Maris.pdf}.
  
\bibitem{StrBlokh}    A. M. Shirokov, V. A. Kulikov, P. Maris,  and J.~P.~Vary,  to be published in 
{\em $NN$ and $3N$ Interactions}, edited by L.~D.~Blokhintsev and
I.~I.~Strakovsky (Nova Science, 2014),
http:/\!/www.novapublishers.com/catalog/ product\verb+_+info.php?products\verb+_+id=49997.
  
  

\bibitem{14F-ex} V. Z. Goldberg, B. T. Roeder, G. V. Rogachev, 
G.~G.~Chubarian, E. D. Johnson, C. Fu, A. A. Alharbi, 
M.~L.~Avila, A. Banu, M. McCleskey, J. P. Mitchell, 
E.~Simmons, G. Tabacaru, L. Trache, and R. E. Tribble, 
 Phys. Lett. B {\bf 692}, 307  (2010).

\bibitem{Bethe} H. A. Bethe, Ann. Rev. Nucl. Sci. {\bf 21}, 93 (1971).

\bibitem{Scott-05} S. K. Bogner, A. Schwenk, R. J. Furnstahl, and A. Nogga,
Nucl. Phys. A {\bf 763}, 59 (2005).

\bibitem{Hebeler-11} K. Hebeler, S.K. Bogner, R.J. Furnstahl, A. Nogga and A. Schwenk, Phys. Rev. C {\bf 83}, 031301 (2011).
\bibitem{Hebeler-13} 
  K.~Hebeler and R.~J.~Furnstahl,
  Phys.\ Rev.\ C {\bf 87}, no. 3, 031302 (2013).
\bibitem{RG1} S. K. Bogner, T. T. S. Kuo, and A. Schwenk, Phys. Rept. 
{\bf 386}, 1 (2003); S. K. Bogner, R. J. Furnstahl and A.~Schwenk, Prog.
Part. Nucl. Phys. {\bf 65}, 94 (2010).

\bibitem{RG2} S. K. Bogner, R. J. Furnstahl, S. Ramanan, and
A.~Schwenk, Nucl. Phys. A {\bf 784}, 79 (2007).

\bibitem{collapse} J. Kuckei, F. Montani, H. M\"uther, and A. Sedrakian,
Nucl. Phys. A {\bf 723}, 32 (2003).

\bibitem{no-collapse-rel} E. N. E. van Dalen and  H. M\"uther, 
Phys. Rev. C {\bf 80},  037303 (2009).

\bibitem{Baardsen} 
  G.~Baardsen, A.~Ekstršm, G.~Hagen and M.~Hjorth-Jensen,
  Phys.\ Rev.\ C {\bf 88}, 054312 (2013).
  
%

\bibitem{highJ70} D. W. L. Sprung, P. K. Banerjee, A. M. Jopko,
and M.~K.~Srivastava, Nucl. Phys. A {\bf 144}, 245 (1970).

\bibitem{highJ79} P. Grange, A. Lejeune, and C. Mahaux, 
Nucl. Phys. A {\bf 319}, 50 (1979).

\bibitem{highJ10} M. Modarres  and T. Pourmirjafari,
Nucl. Phys. A {\bf 848}, 92 (2010).

\bibitem{Nijm-PWA} V. G. J. Stoks, R. A. M. Klomp, M. C. M. Rentmeester, 
and J. J. de Swart, Phys. Rev. C {\bf 48}, 792 (1993); see also
http:/\!/nn-online.org/NN.

\bibitem{Wiringa} R. B. Wiringa, V. Fiks, and A. Fabrocini, Phys. Rev. C {\bf 38}, 1010 (1988).



\bibitem{Akmal} A. Akmal, V. R. Pandharipande, and D. G. Ravenhall, Phys. Rev. C {\bf 58}, 
1804 (1998).

\bibitem{Hagen:2013} 
  G.~Hagen, T.~Papenbrock, A.~Ekstršm, K.~A.~Wendt, G.~Baardsen, S.~Gandolfi, M.~Hjorth-Jensen and C.~J.~Horowitz,
  Phys.\ Rev.\ C {\bf 89}, 014319 (2014).

\bibitem{Gezerlis:2013} 
  A.~Gezerlis, I.~Tews, E.~Epelbaum, S.~Gandolfi, K.~Hebeler, A.~Nogga and A.~Schwenk,
  Phys.\ Rev.\ Lett.\  {\bf 111}, no. 3, 032501 (2013).
  \bibitem{Tews:2013} 
  I.~Tews, T.~KrŸger, A.~Gezerlis, K.~Hebeler and A.~Schwenk,
  arXiv:1310.3643 [nucl-th].


\bibitem{JISP6} A. M. Shirokov, J. P. Vary, A. I. Mazur, S. A. Zaytsev, 
and  T. A. Weber,   Phys. Lett. B  {\bf 621}, 96 (2005). 

\bibitem{JISP6G} A. M. Shirokov, J. P. Vary, A. I. Mazur, S. A. Zaytsev,  
and  T. A. Weber,   J. Phys. G {\bf 31}, S1283 (2005).  
  
 

\bibitem{Bethe:1971} H. A. Bethe, Annu. Rev. Nucl. Sci. {\bf 21}, 93 (1971).



\end{thebibliography}
\end{document}